# Dynamic aspects of individual design activities

## A cognitive ergonomics viewpoint

Willemien VISSER
INRIA - National Research Institute for Computer Science and Control
EIFFEL - Cognition & Cooperation in Design - Rocquencourt
78153 Le Chesnay Cedex (France)
Email: Willemien.Visser@inria.fr

**Abstract**. This paper focuses on the *use* of knowledge possessed by designers. Data collection was based on observations (by the cognitive ergonomics researcher) and simultaneous verbalisations (by the designers) in empirical studies conducted in the context of industrial design projects. The contribution of this research is typical of cognitive ergonomics, in that it provides data on actual activities implemented by designers in their actual work situation (rather than on prescribed and/or idealised processes and methods). Data presented concern global strategies (the way in which designers actually organise their activity) and local strategies (reuse in design). Results from cognitive ergonomics and other research that challenges the way in which people are supposed to work with existing systems are generally not received warmly. Abundant corroboration of such results is required before industry may consider taking them into account. The opportunistic organisation of design activity is taken here as an example of this reluctance. The results concerning this aspect of design have been verified repeatedly, but only prototypes and experimental systems implementing some of the requirements formulated on their basis, are under development.

## 1. Introduction

In a collection of papers presenting the *Developments in design methodology*, Cross (1984) presented "prescription of an ideal design process" as the first of the four stages that he distinguished. This stage, qualified as the period of "systematic design", was reflected by texts dating from 1962 to 1967. Stages two, three and four were "description of the intrinsic nature of design activity" (1966-1973), "observation of the reality of design activity" (late 1970s), and "reflection on the fundamental concepts of design" (1972-1982). Nearly twenty years after 1984, normative, prescriptive models of design (exemplified by Pahl & Beitz 1977, 1984) are still very powerful —even if the *Human Behaviour in Design* Symposium (2003, this book), organised in Pahl and Beitz' Germany with its particularly strong methodological tradition, testifies to a movement that might correspond to the second and third stages identified by Cross (1984): Are they stages of a new cycle?
Few models of actual design have been developed, however, and the proposals that have been made generally focus on particular aspects of the design process (Dorst 1997; Simon 1999; Visser 2002). In order to formulate such models, it is necessary to observe the actual activities carried out by designers. Studies conducted in cognitive psychology and cognitive ergonomics may collect such data. It is this approach to design that is central to this paper which presents a cognitive ergonomics viewpoint on individual design activities.

In the rest of this text, unless specified otherwise, the "design" referred to (through expressions such as "studies on design", or "design is opportunistic") is the actual design activity implemented by designers during their work on design projects —as opposed to normatively based descriptions or prescriptions of design methods or design processes.

**Outset of the paper.** We will describe our approach and illustrate it by data concerning two design strategies adopted by individual designers, i.e. the opportunistic organisation of design, and reuse. In our Conclusion, we will discuss the reluctance of industry to accept results that do not tally with design methods that are to be used in industrial design projects.

## 1.1 Relevance of data concerning individual design activities

Until some ten years ago, most design research was concerned with individual design. In more recent years, an important shift in research on the cognitive aspects of design has consisted in studying, perhaps even to a greater degree, collectively conducted design activities (Blessing, Brassac, Darses, & Visser 2000; Cross, Christiaans, & Dorst 1996). This change in focus, which has also been observed in other task domains, has been accompanied by the acceptance of design as a valid field of research in cognitive ergonomics for which analysis methods are being developed (Darses, Détienne, Falzon, & Visser, 2001). Yet, a majority of studies on design are being conducted in artificially restricted laboratory conditions, in which the design situation is rather different from that in professional design situations (see e.g. most studies in the 1997 Special issue of Design Studies on Descriptive models of design). Nevertheless, compared to other problem solving activities, design has started to be examined rather frequently in actual working situations. Recently, design is even being studied in large-scale industrial settings (Détienne & Falzon 2001; Visser 1993).

The present text is restricted to individual design. The continuing relevance of data concerning individual design can be justified on at least three grounds. First, even if many design projects are undertaken by large teams involving big numbers of designers, engineers and other people; even if discussion, negotiation, cognitive and operative synchronisation play a crucial role in the generation and evaluation of solutions; an important proportion of design activity remains the work of single individuals, especially during distributed-design stages. In addition, even during co-design stages, cognitive activities in collective design are those implemented in individual design to which are added activities that are specific to co-operative work. We see no evidence to suppose that co-operation modifies the nature of the elementary problem solving processes implemented in design, i.e. solution development and evaluation processes (Visser 1993). Finally, the development of appropriate work environments for designers, such as shared and private work spaces in computer-mediated design, requires the analysis of the links between the different forms of reasoning implemented in both individually and collectively conducted activities.

## 1.2 Dynamic aspects of individual design activities

Within the domain of individual design, this paper focuses on its dynamic aspects. Examples will primarily come from our own work. For some 20 years now, we have been collecting data on the use of knowledge, generally via design strategies. We did so through empirical studies, often conducted in the context of industrial design projects in different application domains, mainly software development, mechanical and industrial design.

The focus on dynamic aspects of design activities, i.e. on the *use* of knowledge, requires some comment. Many studies do indeed concern knowledge that is used in design, or knowledge that may be used in design. However, these studies seldom concern the way in which this knowledge is actually used, i.e. the modalities and conditions of their use in design activities. In one of our first methodological studies on software design, we compared these two aspects of design knowledge use (Visser 1985, see also Visser & Morais 1991). We confronted two types of data that we had collected using different data collection methods. Four of the methods proposed (interviews and analyses of the result of the activity) may reveal knowledge possessed by designers, but it remains hypothetical *if* and *how* they use it. Only one of the methods (observing designers during their activity) may provide data on knowledge possessed by designers *and* on its actual use. The observational method is expensive and can usually only be applied to a few designers. It thus requires, in general, independent validation of its results.

## 2. Actually implemented design strategies

The "systematic design" movement in industrial, engineering and architectural design has its counterpart in software design, in the form of the "waterfall" model, and other "structured" and "stepwise refinement" methods. Early empirical studies conducted on design, especially in the domain of software design, generally characterise designers' activity as following such methods, i.e. as well structured and even as hierarchically organised, in other words, as following a pre-established plan. They assert that designers' global control strategy consists in decomposing their problem according to a combined top-down, breadth-first strategy. Both are general problem solving strategies that are not specific to design, but can be used in nearly any problem-solving task. These are the strategies identified, analysed and detailed in classical problem solving research, from Newell and Simon (1972) on.

In these early design studies, a trend seen was to "conflate prescriptive and descriptive remarks" on the activity (Carroll & Rosson 1985). Rather than to consider what the activity was really like, researchers focused on what it should be (Visser & Hoc, 1990). One example is the study by Jeffries, Turner, Polson and Atwood (1981). According to these authors, "a reasonable model of performance... ought to be related to accepted standards of good practice" and "most expert designers are familiar with this literature and may incorporate facets of these methodologies into their designs" (p. 256).

More recent studies, however, observing designers in realistic situations or even in real work situations, show that the strategies implemented by these designers deviate from the top-down and breadth-first prescriptive model, and lead rather to an opportunistically organised design activity. In our own studies on software design we observed top-down and breadth-first decomposition strategies to be implemented, but only locally. Their combination did not seem to be the control strategy of the design activity at the global, organisational level. Other strategies were implemented at a local level. These could be strategies already identified in the problem solving research literature, e.g. simulation, or strategies that had not yet been presented in the literature, e.g. reuse.

The two examples of strategies actually implemented by professional designers when working on their design projects and presented in this paper, are the global strategy used by designers in order to organise their activity, i.e. the opportunistic-organisation strategy, and a more locally applied strategy, i.e. reuse.

**Data collection.** Data collection in our studies referred to below was based on observation (by the cognitive ergonomics researcher) and simultaneous verbalisation (by the designers).

This approach is typical of cognitive ergonomics research, in that it provides data on the activities implemented by designers in their actual work situation (rather than on prescribed and/or idealised processes, or on actual activities observed in artificially restricted conditions).

## 2.1 Opportunistic organisation of design

Already in 1980, Green (1980) concludes a discussion of structured programming methods by advancing the idea that "good programmers…. leap intuitively ahead, from stepping stone to stepping stone, following a vision of the final program; and then they solidify, check, and construct a proper path. That proper path from first move to last move, in the correct order, is the program, their equivalent of the formal proof." (p. 306) Green notes that Wirth, who introduced the concept of "stepwise refinement", is himself "quite explicit; having described his stepwise refinement, [Wirth] says 'I should like to stress that we should not be led to infer that the actual program development proceeds in such a well organised, straightforward, top-down manner. Later refinement steps may show that earlier ones are inappropriate and must be reconsidered.'" (Ibid.)

The qualification that empirical studies on actual design activities use for the way in which designers organise their activity is "opportunistic" (Visser 1988a). Analysing as a design activity, the errand planning modelled by Hayes-Roth and Hayes-Roth (1979) as an opportunistically organised activity, we followed these authors' approach, and qualified the organisation of design activity as "opportunistic". We did so because designers' selection of consecutive design actions is determined by an evaluation function that is primarily based on the "cognitive economy" criterion, rather than by a pre-established plan, be it hierarchical or otherwise. Such plans may play a role, and often their role will be important, but they are only one of several possible resources that provide opportunities for action.

We observed this opportunistic design in different domains, e.g. in a series of studies conducted on three consecutive stages of an industrial design project: mechanical designers defining the functional specification for programmable-controller (PC) software design (Visser 1988a); a software engineer designing the PC software (Visser 1987); and a team composed of a software engineer and mechanical designers "testing" the PC software (but in fact also redesigning it) (Visser 1988b). In these studies, we identified six categories of data that could be "taken advantage of" as factors leading to the opportunistic organisation of design.

In this paper, only one category will be presented through an example from the observations made on `the functional-specifications design (for the other categories and examples, see Visser 1994). Designers can, for example, take advantage of mental representations of a design object related to the representations that they are using for their current design action. Analogy is one example of such a relationship. Considering second-phase tooling operations (in order to finish the rods) as analogous to the first-phase tooling operations (in order to shape the rods) on which he is currently working, the designer continually switches between their specifications. Often he takes advantage of the specification of a first-phase tooling operation O1 to specify, adapting this O1 specification, its corresponding second-phase tooling operation O2. Frequently, an O1 specification "makes him think" of an omission or error made on the corresponding O2 operation.

The observation that designers organise their activity in an opportunistic way is not restricted to inexperienced designers. On the contrary, it is something typical of expert designers. Nor is it the translation of a deteriorated behaviour that occurs only when designers are confronted with a "difficult" design task. Even when expert designers are involved in routine tasks, the retrieval of pre-existing plans does not appropriately characterise the organisation of their actual activity. An analysis of 15 empirical studies on design (Visser 1994) showed that
- even if designers possess a pre-existing solution plan for a design problem,
- and if they can and, in fact, do retrieve this plan to solve their problem (which is often possible for expert designers confronted with routine design),
- yet if other possibilities for action ("opportunities") are also perceived (which is often the case in real design)
- and if the designers evaluate the cost of all possible actions ("cognitive" and other costs), as they will do in real design,
- the action selected for execution will often be an action other than the one proposed by the plan: it will indeed be a selected opportunity.

Pre-existing plans that —if they are invariably followed— may lead to systematically organised activities, are only one of the various action-proposing knowledge structures used by designers. They may be interesting from a cognitive-economy viewpoint because executing an action for which such a schematic memory representation is available may cost relatively little if all schema variables relevant for execution have constant or default values. But if other knowledge structures propose relatively more economical actions, designers may deviate from such a plan. This is especially true for experts, who may be supposed to possess —or else to be able to construct without difficulty— a representation of their activity which allows them to resume their plan later on, when it once again becomes profitable to do so from the viewpoint of cognitive economy. Having to compare several action proposals and taking into account the cognitive cost of an action are two task characteristics which probably only appear in "real" design. This may explain why in laboratory experiments mostly systematically organised design activities were observed.

"Opportunities" must be "perceived": this perception is data-driven. It is on the basis of their knowledge that expert designers process the data that they perceive in their environment and that may take different forms: information (the state of design in progress, but also other information at the designers' disposal, information they receive, information they construct), permanent knowledge and temporary design representations (in particular, the designers' representations of the state of design). Taking advantage of these "opportunities" rather than following a pre-established plan will, indeed, lead to an opportunistically organised activity.

## *2.2 Reuse*

All use of knowledge could be called "reuse" in that knowledge is based on the processing of previous experience and data encountered in the past. We reserve "reuse" (vs. other "use" of knowledge) for the use of specific knowledge (the "source" knowledge) that is at the same abstraction level as the "target" (the design problem to be solved) for whose processing the knowledge in question is retrieved. Thus, reuse of knowledge is opposed to the use of more general, abstract knowledge (such as knowledge structures like schemas and rules).

Reuse has been identified in various empirical design studies. The exploitation of specific experiences from the past is indeed particularly useful in design, especially in non-routine design (Visser 1993, 1995a).

Software-engineering researchers distinguish "design for reuse" from "reuse for design". The construction of reusable entities that are to be organised into a "components library" is often considered an independent design task, not necessarily executed by the designer who is going to reuse these entities. We are unaware, however, of any empirical studies conducted in such "design for reuse" situations. Existing empirical studies concern "reuse for design" and show that the two activities are not as separate as software-engineering researchers suppose.

A considerable proportion of the empirical, cognitive ergonomic research on reuse in design has been conducted in the domain of software, especially that of object-oriented (OO) software (Détienne 2002). Visser (1987) observed reuse in programmable controller-software design using a declarative type of language. She also studied reuse in other domains, i.e. mechanical (Visser 1991) and industrial design (Visser 1995b).

Several aspects of reuse have been examined in these studies. In this paper, we only discuss the question that is at the basis of reuse-based design, i.e.: When do designers decide to adopt reuse in order to solve a design problem, rather than base their problem solving on general knowledge, i.e. proceed to "design from scratch"?

Reuse takes place in, at least, five stages[1]: 0. construction of a representation of the target problem; 1. retrieval of one or more sources; 2. adaptation of the source into a target-solution proposal; 3. evaluation of the target-solution proposal; 4. integration into memory of the resulting modifications in problem and solution representations.

The construction of a target problem representation (stage 0) has seldom received attention in empirical studies. It is, however, during this stage that designers have decide whether they are going to (try to) adopt reuse in order to solve their design problem. We are not aware of any study informing us about designers taking into consideration the choice between design from scratch and reuse —rather than to "simply" design from scratch right from the start. Two individual studies seem to indicate that the "cost" of reuse is the main factor in the decision process whether to proceed to reuse (Burkhardt & Détienne 1995; Visser 1987).

In an experimental study, Burkhardt asked seven OO-software designers to describe elements they might want to (re)use. Half of his subjects mention that there are reusable elements whose actual reuse they would not envision because of the "cost" of their reuse. Data gathered by Visser (1987) present an example of a factor contributing to the cost of reuse —but only once candidate sources, i.e. reusable solutions, have been retrieved. This factor is the cost of required adaptation, itself a function of target-source similarity. Sources are indeed always to be adapted, in order to be usable as a possible solution to a target problem that is "similar" to the source problem that had been solved by the source. This conclusion coincides with a position adopted in several case-based reasoning (CBR) systems in which the selection of a "case", i.e. a reusable source, is guided by its adaptability (Smyth & Keane 1995).

The importance of the "cost" factor in the choice of a strategy such as reuse is completely in line with our identification of the primordial factor underlying the organisation of design, i.e. cognitive economy (Visser 1994). It is the relative cost of an action that determines its choice —if a designer is conscious of several possibilities.

With respect to the frequency of reuse, different authors in the domain of software engineering assert that 40 - 80 % of code is non-specific, thus reusable. Many authors advance the percentage of 80 %, mostly referring to Jones (1984), who summarises four studies conducted between 1977 and 1983. In 1989, however, Biggerstaff and Perlis assert that "over the broad span of systems, reuse is exploited today but to a very limited extent".

Empirical studies providing quantitative data all concern OO software, which, due to its mechanisms of inheritance, abstraction and encapsulation, and polymorphism, is considered to particularly "favour reuse", so the conclusions of these studies may be specific to this form of software design. As far as we know, the only empirical study observing "massive" reuse concerns OO software (Lange and Moher 1989). If empirical studies on other design tasks do not provide data on the frequency of reuse, experimental studies on analogical reasoning can inform us. Their general conclusion is that source retrieval seldom occurs "spontaneously", i.e. without having been suggested by the experimenter —which often occurs in these studies. The data available on reuse-based design vs. design from scratch is thus still very meagre. Nevertheless, it is a central topic with respect to reuse, its role in design, and the possibilities to support designers in their reuse during design.

---

[1] Our use of the term "stage" does not mean that such stages are purely consecutive, and that a previous stage cannot be returned to subsequently.

## 3. Conclusion

Our results concerning the organisation of design may inspire, at least, two approaches to design assistance: given the opportunistic organisation of the activity of designers working "in total freedom", one may either try to use tools to prevent such a way of proceeding, or offer designers tools that would assist them in their "natural" way of proceeding. As a researcher in cognitive ergonomics, we consider that assistance tools should be compatible with the actual activity, i.e. with the designers' mental structures and processes. If designers' activities are opportunistically organised because of reasons of cognitively interesting management, a support system which supposes —and therefore imposes— a hierarchically structured design process will at the very least constrain designers and may well even handicap them. Most tools continue, however, to be based on prescriptive or analytical, task-based models of design, i.e. not on data concerning the actual activity that is to be supported.

Results from cognitive ergonomics and other research that challenge the way in which people are supposed to work with existing systems are generally not received warmly. Abundant corroboration of such results is required before industry may consider taking them into account. The opportunistic organisation of design activity can be taken as one example of this reluctance. The results concerning this aspect of design have been verified repeatedly now, requirements for systems offering designers "real" support have been formulated on the basis of these results (see e.g. Visser & Hoc 1990), but only prototypes and experimental systems implementing some of these requirements are under development. An example of a software design environment based on the results concerning opportunism is GOOSE (Generalised Object Oriented Support Environment), the experimental CASE (Computer Assisted Software Engineering) tool for OO software design developed at the University of Keele by David Budgen and colleagues. One of the specifications for GOOSE was that it should enable its users to adopt an opportunistic strategy in developing their ideas. However, no commercial tools integrating these elements are available for use in industry, i.e. in "real" design, which is the focus of this paper.

Certain findings presented in this text may seem "obvious": one might think that any sensible person, especially a designer, might formulate them "simply" on the basis of their experience, or even of their common sense. One might think that no empirical studies are required to obtain the knowledge corresponding to these results. Studies in the domain of cognitive psychology and ergonomics teach us, however, that such "common-sense" judgements concerning "obvious" phenomena are fallacious, even if they correspond to the intuition of people who are proficient in the domain. Contrary to a widespread opinion, the fact that designers deviate from, or even do not follow at all, procedures that are prescribed by the majority of design methods, is not due to designers being nonchalant, making errors or displaying other deficiencies. These deviations or even abandoning of imposed structures have cognitive causes that are worth being examined and especially taken seriously in the development of design environments or other support modalities for designers.

Precisely because of their grounding in empirical studies conducted in real, complex situations, following the research methodology of cognitive psychology, these data are valuable as a basis for such system development. They constitute a basis that is more valuable than, on the one hand, psychological research conducted in restricted laboratory conditions, or, on the other hand, models based on introspection, norms, or other prescription-based models.

# References

Biggerstaff, T. J., & Perlis, A. J. (1989). Introduction. In T. J. Biggerstaff & A. J. Perlis (Eds.), Software reusability (Vol. 1). Reading, MA: Addison-Wesley.

Blessing, L., Brassac, C., Darses, F., & Visser, W. (Eds.). (2000). Analysing and modelling collective design activities. Proceedings of COOP 2000, Fourth International Conference on the Design of Cooperative Systems. Rocquencourt: INRIA.

Burkhardt, J.-M., & Détienne, F. (1995, 27-29 June 1995). An empirical study of software reuse by experts in object-oriented design. Proceedings of Interact'95, Lillehammer (Norway).

Carroll, J. M., & Rosson, M. B. (1985). Usability specifications as a tool in iterative development. In H. R. Hartson (Ed.), Advances in human-computer interaction. Norwood, NJ: Ablex.

Cross, N. (1984). Developments in design methodology: Wiley.

Cross, N., Christiaans, H., & Dorst, K. (Eds.). (1996). Analysing design activity. Chichester: Wiley.

Darses, F., Détienne, F., Falzon, P., & Visser, W. (2001). COMET: A method for analysing collective design processes (Rapport de Recherche INRIA 4258). Rocquencourt: INRIA.

Design Studies. (1997). Special issue on Descriptive models of design. Design Studies, 18(4).

Détienne, F. (2002). Software design. Cognitive aspects. London: Springer.

Détienne, F., & Falzon, P. (2001). Cognition and Cooperation in Design: the Eiffel research group. In M. Hirose (Ed.), Human-Computer Interaction-Interact 2001 (pp. 879-880).

Dorst, K. (1997). Describing design. A comparison of paradigms. Delft: TU Delft

Green, T. R. G. (1980). Programming as a cognitive activity. In H. T. Smith & T. R. G. Green (Eds.), Human interaction with computers. London: Academic Press.

Hayes-Roth, B., & Hayes-Roth, F. (1979). A cognitive model of planning. Cognitive Science, 3, 275-310.

Jeffries, R., Turner, A. A., Polson, P. G., & Atwood, M. E. (1981). The processes involved in designing software. In J. R. Anderson (Ed.), Cognitive skills and their acquisition. Hillsdale, NJ.: Erlbaum.

Lange, B. M., & Moher, T. (1989). Some strategies for reuse in an object-oriented programming environment. Proceedings of CHI'89, Austin (USA).

Newell, A., & Simon, H. A. (1972). Human problem solving: Prentice Hall.

Pahl, G., & Beitz, W. (1977). Konstruktionslehre. Berlin: Springer.

Pahl, G., & Beitz, W. (1984). Engineering design. London: The Design Council.

Simon, H. A. (1999). The sciences of the artificial (3rd, rev. ed. 1996; Orig. ed. 1969). Cambridge, MA: The MIT Press.

Smyth, B., & Keane, M. T. (1995). Some experiments on adaptation-guided retrieval. In M. Veloso & A. Aamodt (Eds.), Case-based reasoning (pp. 313-324). Berlin: Springer.

Visser, W. (1985). Modélisation de l'activité de programmation de systèmes de commande

[Modelling the activityof programming control systems] (In French). Actes du colloque COGNITIVA 85 (Tome 2), Paris.

Visser, W. (1987). Strategies in programming programmable controllers: a field study on a professional programmer. In G. Olson, S. Sheppard, & E. Soloway (Eds.), Empirical Studies of programmers: Second Workshop (pp. 217-230). Norwood, NJ: Ablex.

Visser, W. (1988a). Giving up a hierarchical plan in a design activity (Rapport de recherche 814). Rocquencourt, France: INRIA.

Visser, W. (1988b). L'activité de comparaison de représentations dans la mise au point de programmes. Le Travail Humain, Numéro Spécial "Psychologie ergonomique de la programmation informatique", 51(4), 351-362.

Visser, W. (1991). Evocation and elaboration of solutions: Different types of problem-solving actions. An empirical study on the design of an aerospace artifact. COGNITIVA 90, Paris.

Visser, W. (1993). Collective design: A cognitive analysis of cooperation in practice. ICED 93, the 9th International Conference on Engineering Design, The Hague (The Netherlands).

Visser, W. (1994). Organisation of design activities: opportunistic, with hierarchical episodes. Interacting With Computers, 6(3), 239-274 (Executive summary: 235-238).

Visser, W. (1995a). Reuse of knowledge: empirical studies. In M. Veloso & A. Aamodt (Eds.), Case-based reasoning. Berlin: Springer.

Visser, W. (1995b). Use of episodic knowledge and information in design problem solving. Design Studies, 16(2), 171-187.

Visser, W. (2002). A Tribute to Simon, and some —too late— questions, by a cognitive ergonomist. International Conference In Honour of Herbert Simon "The Sciences of Design. The Scientific Challenge for the 21st Century", Lyon (France), INSA.

Visser, W., & Hoc, J.M. (1990). Expert software design strategies. In J.M. Hoc, T. Green, R. Samurçay, & D. Gilmore (Eds.), Psychology of programming (pp. 235-250). London: Academic Press.

Visser, W., & Morais, A. (1991). Concurrent use of different expertise elicitation methods applied to the study of the programming activity. In M. J. Tauber & D. Ackermann (Eds.), Mental models and human-computer interaction. Amsterdam: Elsevier.